\documentclass[twocolumn,pra,preprintnumbers,superscriptaddress]{revtex4-2}

\usepackage{graphicx, dcolumn, bm, amsmath, dsfont, xcolor, comment, physics}

\usepackage{titletoc}

\makeatletter
\def\maketitle{
\@author@finish
\title@column\titleblock@produce
\suppressfloats[t]}
\makeatother

\begin{document}

\title{Entanglement percolation enhanced by the power of choice}

\author{Yanxuan Shao}
\affiliation{Department of Physics and Astronomy, Northwestern University, Evanston, Illinois 60208, USA}
\affiliation{Center for Network Dynamics, Northwestern University, Evanston, Illinois 60208, USA}

\author{Adilson E. Motter}
\affiliation{Department of Physics and Astronomy, Northwestern University, Evanston, Illinois 60208, USA}
\affiliation{Center for Network Dynamics, Northwestern University, Evanston, Illinois 60208, USA}
\affiliation{Department of Engineering Sciences and Applied Mathematics, Northwestern University, Evanston, Illinois 60208, USA}
\affiliation{Northwestern Institute on Complex Systems, Northwestern University, Evanston, Illinois 60208, USA}
\affiliation{Institute for Quantum Information Research and Engineering, Northwestern University, Evanston, Illinois 60208, USA}

\begin{abstract}
Quantum communication networks are best supported by maximally entangled states, but imperfections 
in entanglement generation and distribution generally 
limit the distributed states to being only partially entangled.
Therefore, protocols typically consisting of 
local operations and classical communication
are required to create suitably entangled networks. Here, we introduce a power-of-choice (PoC) protocol based on reversing the selection criteria in the Achlioptas process previously considered in classical explosive percolation. The resulting PoC protocol anticipates the entanglement percolation transition, thereby reducing the entanglement resources needed to establish quantum communication. Notably, in contrast to explosive percolation, the PoC protocol admits an implementation using local information with no appreciable loss of performance. 
\end{abstract}

\maketitle

\section{Introduction}
Entanglement-based networks enable quantum communication between remote nodes \cite{hasan2023quantum, bassoli2021quantum}, laying a foundation for distributed quantum information applications such as networked quantum sensing \cite{kim2024distributed, zhang2021distributed, zhao2021field}, distributed computing \cite{main2025distributed, caleffi2024distributed, gyongyosi2021scalable}, and a possible quantum internet \cite{kumar2025quantum, wehner2018quantum}.
However, establishing long-distance entanglement remains challenging, and many efforts are underway to address this limitation, including the development of quantum repeaters \cite{azuma2023from, wallnofer2024faithfully, briegel1998quantum},  use of alternative media for entanglement distribution \cite{thomas2024quantum, bulla2023nonlocal, yin2017satellite}, and design of new protocols for entanglement generation and distribution \cite{xiao2024connectionless, patil2022entanglement}.
In addition to distance-related losses,
one must also contend with imperfections and decoherence in entanglement generation and distribution, resulting in only partially entangled or mixed states being available in practice \cite{lei2023quantum, newmann2022continuous}. Given a network-wide distribution of such states, local quantum operations and classical communication can be used to manipulate entanglement to establish high end-to-end fidelity. This naturally leads to the problem of quantum entanglement percolation \cite{perseguers2013distribution, Nokkala2025}, where the central question is how to generate a connected network of high-fidelity states starting from lower fidelity ones using free operations with respect to the entanglement resources.

The canonical form of quantum entanglement percolation begins with a network of pure nonmaximally entangled pairs and uses singlet conversion to create a percolating cluster of maximally entangled edges. 
The resulting cluster enables Bell-state connectivity between arbitrary intracluster nodes via entanglement swapping \cite{pan1998experimental}.
Pioneering work 
established a connection between entanglement percolation and classical percolation theory in statistical physics \cite{acin2007entanglement}. Subsequent work has considered protocols that can reduce the entanglement budget
for percolation for a wide range of conditions.
For example, it has been shown that the percolation threshold can be lowered in arbitrary random networks by preceding the singlet conversion with entanglement swapping to change the network topology \cite{cuquet2009entanglement},
by performing multipartite measurements on bipartite states to form Greenberger-Horne-Zeilinger states in various lattices \cite{perseguers2010multipartite}, and by performing entanglement distillation and swapping before singlet conversion 
\cite{meng2021concurrence, malik2022concurrence}.
These and other prior works have focused mainly on 
promoting percolation by combining entanglement-manipulation operations.
A central open problem is whether entanglement percolation can be further enhanced through singlet-conversion strategies alone, while other local operations and classical communication (LOCC) techniques may still provide additional gains.

We approach this problem inspired by the Achlioptas process, originally introduced as an alternative to traditional bond percolation that can generate
explosive percolation transitions in classical networks \cite{dsouza2015anomalous, achlioptas2009explosive}. In this process, two edges, rather than one, are selected at each time step, and then one is realized, namely, the one that delays the most the formation of a large connected cluster. The resulting delayed percolation transition is abrupt and can be continuous \cite{riordan2011explosive, grassberger2011explosive, lee2011continuity, dacosta2010explosive}  or discontinuous \cite{dsouza2019explosive, cho2013avoiding, nagler2012continuous, boettcher2012ordinary}, depending on the substrate network and specific 
edge selection criteria---a degree of freedom that can be referred to as the {\it power of choice} \cite{dsouza2015anomalous}.

In this article, we introduce a power-of-choice protocol for entanglement percolation that anticipates the percolation transition, which is based on reversing the edge selection criteria used in explosive percolation. 
This allows us to switch from the classical entanglement-budget problem to determining the minimum number of singlet conversions, counting both successful and unsuccessful ones, required to percolate.
For maximally entangled states generated through singlet conversion,
our protocol accelerates the emergence of a percolating network of Bell states for the entire range of singlet-conversion probabilities. 
Rate-equation analysis reveals that the percolation transition remains continuous with a characteristic power-law distribution of cluster sizes. Most importantly, an implementation of the protocol that
circumvents its reliance on cluster-size comparisons is shown to exhibit essentially the same percolation curve as the implementation that uses global information. We suggest that this surprising result can facilitate decentralized large-scale applications of relevance to a future quantum internet.

\begin{figure}[t]
    \centering
    \includegraphics[width=0.43\textwidth]{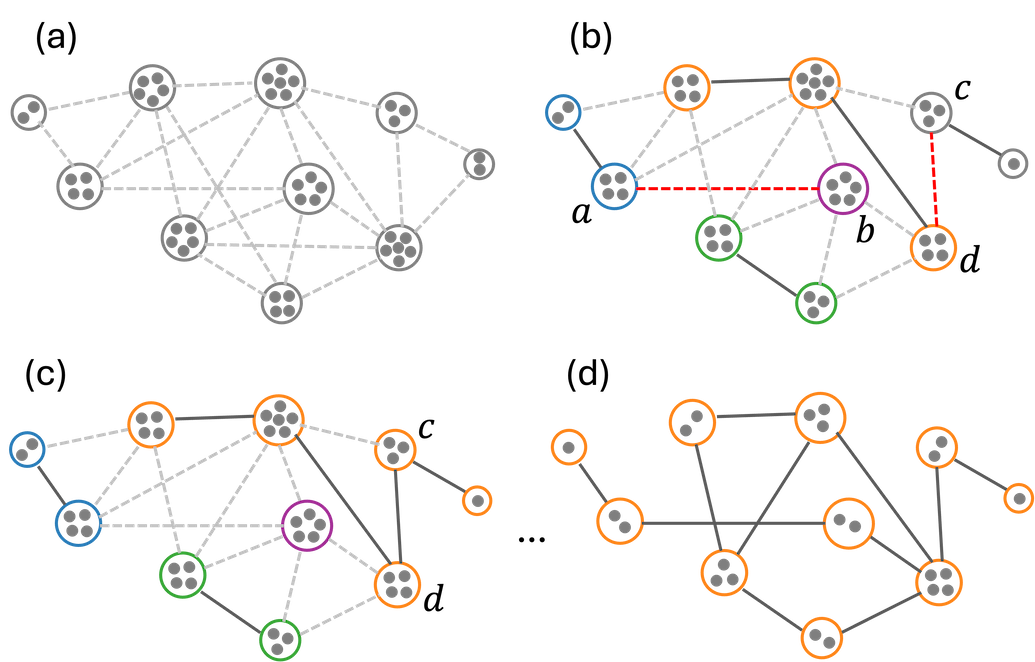}
    \caption{Schematic diagram of the PoC protocol on a quantum network. The nodes (circles) possess one or more qubits (dots) through which entanglement is established with other nodes.
    (a) An initial network of partially entangled states (dashed lines). 
    (b) A step in the percolation process, where Bell states obtained from previous singlet conversions (solid lines) connect nodes into clusters (color-coded nodes). When the two red edges are sampled, edge $(c,d)$  is chosen for singlet conversion because the cluster it would form is larger. (c)~Successful singlet conversion of the edge selected in (b). 
    (d)~A realization of the network of Bell states that results from iterating the process in (b) and (c).}
    \label{fig:schematic}
\end{figure}

\section{Power-of-choice protocol} 
We consider quantum networks consisting of $N$ quantum nodes, each equipped with quantum memories storing one or more qubits. 
Initially, each node is connected, on average, to $\bar{k}$ different {\it nodes}, with each connection ({\it edge}) consisting of a nonmaximally entangled two-qubit state that is assumed, for simplicity, to be pure, identical across all edges, and free of decoherence throughout the percolation process.
Up to a change of basis, such pairwise entangled states can be expressed as $\ket{\phi}=\sqrt{\lambda_0}\ket{00}+\sqrt{\lambda_1}\ket{11}$, where $\lambda_0$ and $\lambda_1$ are Schmidt coefficients satisfying $\lambda_0+\lambda_1=1$ and $\lambda_0\geq \lambda_1>0$ ($\lambda_0=\lambda_1=\frac{1}{2}$ in the case of a Bell state) \cite{nielsen2010quantum}. Using only LOCC,
each partial entanglement can be converted into a maximally entangled state with a singlet-conversion probability $p=2\lambda_1\in (0,1)$ \cite{vidal1999entanglement}; with probability $1-p$, the conversion fails and
entanglement is lost.

The entanglement percolation problem can then be framed as the task of performing singlet conversions until a giant connected component is formed. That is, the process continues until a connected cluster containing a finite fraction of all nodes emerges in the limit of large $N$. Once this percolation transition is reached, one may proceed further, possibly until the majority or all nodes belong to a single connected cluster. In either case, the significance of the process is that nodes in the same connected cluster can, in principle, establish direct pairwise entanglement for quantum communication via entanglement swapping, with the understanding that an entire path is consumed by each pair \cite{meng2025path, kim2024shortest-path}.

The traditional approach to entanglement percolation is the {\it random protocol}, in which at each step an unconverted edge is chosen uniformly at random for singlet conversion. As a result of recurrent singlet conversions, the normalized size $S$ of the largest connected cluster, measured by the fraction of the number of nodes it contains, increases from zero starting at the percolation transition point. This process can be mapped to a classical bond percolation problem where the probability of successfully establishing a chosen edge is $p$.

Departing from the random strategy, we introduce a {\it power-of-choice (PoC) protocol} inspired by the Achlioptas process in classical networks \cite{dsouza2019explosive}. At each time step, two distinct nonmaximally entangled edges, $(a,b)$ and $(c,d)$, are randomly selected, and one of them is chosen for singlet conversion according to the following rules:
(1) If either pair of nodes already lies in the same (connected) cluster of the maximally entangled network, that edge is discarded, and the other is selected; if both pairs are intracluster, one is selected at random.
(2) If neither pair lies in the same cluster, the sum of the cluster sizes of nodes $a$ and $b$ is compared with that of $c$ and $d$, and the edge with the larger sum is selected; if the sums are equal, one of the two edges is chosen at random.
Singlet conversion is then performed on the selected edge, and the process is iterated as illustrated in Fig.~\ref{fig:schematic}.
Prior to the percolation transition, the protocol reduces to the second rule without the need to check cluster identity, since the probability that randomly selected edges belong to the same cluster is negligible.

Note that as $S$ increases, a larger fraction of the remaining partially entangled edges lies within the largest connected cluster, and the PoC protocol disfavors singlet conversion on such redundant edges. The protocol can also be generalized by considering $\ell>2$ edges at a time, always preferentially selecting the one that connects different clusters with the largest combined size. In the simulations, cluster merges are tracked using the union-find data structure \cite{newman2001fast}.
We illustrate our main results using Erd\H{o}s-R\'enyi (ER) networks of partial entanglements, generated by connecting each pair of nodes independently with probability $\bar{k}/(N-1)$, so that the average number of connections per node is $\bar{k}$. We focus on the summation rule, which facilitates the rate-equation analysis below, but using the cluster-size product would produce similar percolation curves (Supplemental Material (SM) \cite{SM}, 
Sec.~S1).

\begin{figure}[t]
    \centering
    \includegraphics[width=0.99\columnwidth]{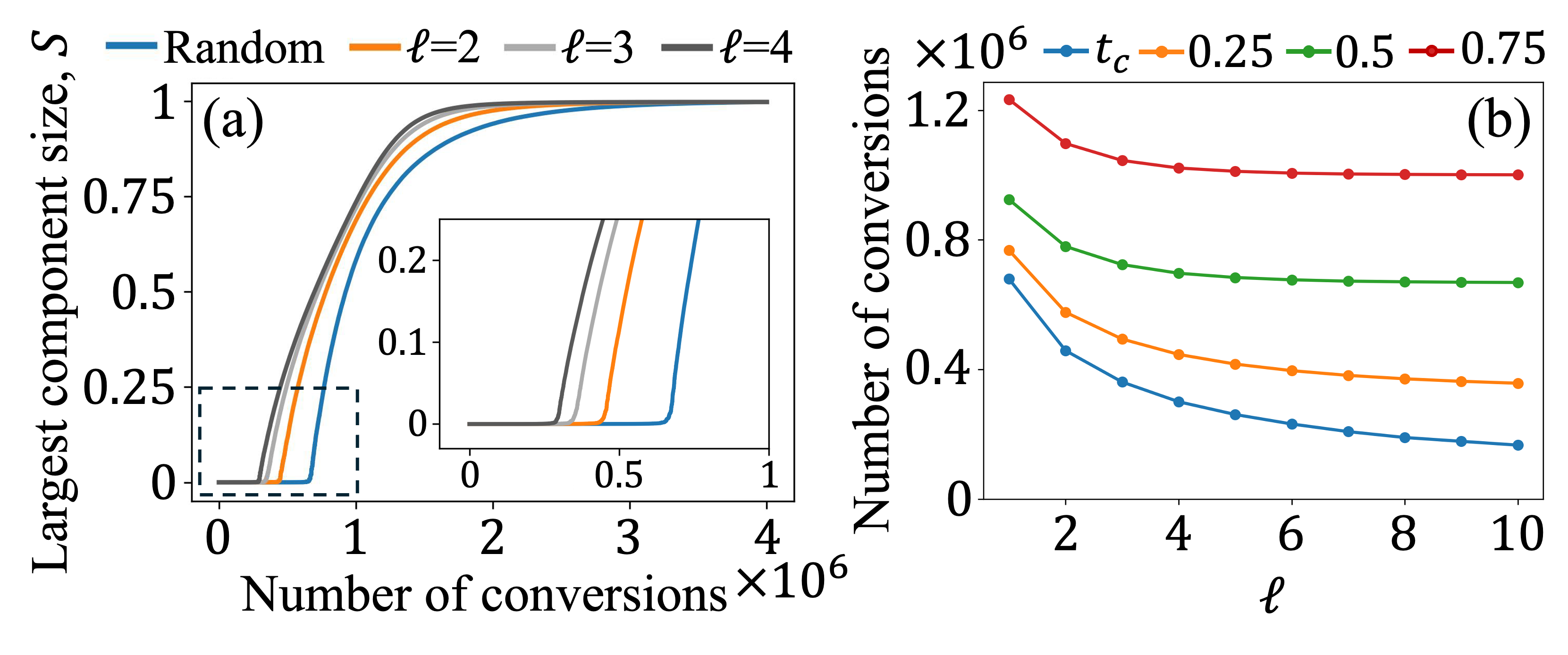}
    \caption{Entanglement percolation for the PoC protocol. (a) Normalized size of the largest connected cluster as a function of the number of singlet conversions for different $\ell$, with the random protocol shown as a reference. Inset: Zoom-in around the percolation transition (dashed region).
    (b) Number of singlet conversions at the percolation threshold $t_c$ (defined below), as well as at the points where the normalized size of the giant connected component reaches 0.25, 0.5, and 0.75, for different values of $\ell$.
    In all cases, the initially prepared partial entanglement networks are ER networks with $N=10^6$ nodes, average degree $\bar{k}=8$, and $p=0.75$. Each curve is an average over ten independent realizations. 
    }
    \label{fig:PoC}
\end{figure}

Figure~\ref{fig:PoC}(a) shows the time evolution of $S$ in networks with $N=10^6$ nodes for the PoC protocol with $\ell=2$, $3$, and $4$ versus the random strategy. 
In all cases, $S$  vanishes below the transition, then grows and saturates at a common value set by $p$ and $\bar{k}$. However, the PoC protocol induces an earlier percolation transition and shifts the entire percolation curve, thereby enhancing the supercritical regime that follows and requiring fewer singlet conversions to achieve a given $S$. Moreover, the larger the number $\ell$ of edges considered at each time step, the earlier percolation occurs, although the extent of this effect diminishes as $\ell$ becomes larger [Fig.~\ref{fig:PoC}(b)].
Extensions of the PoC protocol are introduced below, but first we examine the nature of this percolation transition, focusing on $\ell=2$ (which is also assumed throughout the rest of the paper).

\section{Percolation transition properties}
We use the rate equation method \cite{krapivsky2010kinetic} to 
model the evolution of cluster sizes in the percolation process.  Let $x(s,t)$ be the fraction of nodes that are in clusters of size $s$, where $t$ is the normalized time step, namely, the number of singlet conversions relative to the number of nodes.
Given a protocol, we use $y(s,t)$ to denote the probability that the edge selected for singlet conversion has a node in a cluster of size $s$. 
In the large $N$ limit, the variable $t$ can be treated as continuous, and the evolution of different cluster sizes in ER networks is described by \cite{dacosta2015inverting}
\begin{equation}
    \frac{dx(s,t)}{dt}=p  \left[-2sy(s,t) + s\sum_{u+v=s} y(u,t)y(v,t)
    \right].
    \label{rate_equation}
\end{equation}
The first term in brackets corresponds to the case where one endpoint of the selected edge is originally in a cluster of size $s$, so $x(s,t)$ decreases after a successful singlet conversion. The second term accounts for the case where the selected edge connects two smaller clusters whose total size is $s$, thereby increasing $x(s,t)$ upon successful singlet conversion. 

The relative size of the largest maximally entangled cluster near the percolation transition is
\begin{equation}
    S(t)=1-\sum_{s\leq M}x(s,t), 
    \label{cluster_size}
\end{equation} 
where $M$ is the threshold above which the network is expected to have only one connected cluster \cite{dsouza2010local, spencer2007birth}. Moreover, close to the transition, the largest cluster is expected to be much larger than the second largest one. 
Focusing on ER networks unless noted otherwise, we can now compare the random and PoC protocols.

For the random protocol $y(s,t)= x(s,t)$, and Eq.~\eqref{rate_equation} can be solved recursively from $s=1, 2,...$ 
yielding the well-known solution for standard percolation \cite{bennaim2005kinetic}: 
\begin{equation}
    x(s,t)=\frac{s^{s-1}}{s!}(2p t)^{s-1}e^{-2spt}
    \label{random_protocol}. 
\end{equation}
From Eqs.~\eqref{cluster_size} and \eqref{random_protocol}, 
we obtain
the giant component size $S$ as the nonzero solution to $1-S=e^{-2ptS}$, which exists only if $t>(2p)^{-1}$. Calculating the derivative $\frac{dS}{dt}$ at the percolation transition point $t_\text{c}=\frac{1}{2p}$, one obtains
\begin{equation}
    \frac{dS}{dt}\Big|_{t=(1/2p)^-} = 0, \,\,\, \frac{dS}{dt}\Big|_{t=(1/2p)^+} = 4p>0. 
\end{equation}
The discontinuity of the derivative shows that the random singlet-conversion protocol gives rise, as expected, to a second-order percolation transition.

For the PoC protocol,
$y(s,t)$ is not equivalent to $x(s,t)$, and the nature of the transition is not immediate. To calculate $y(s,t)$, we note that for large $N$, the fraction of redundant edges is negligible before a giant connected component emerges.
Near percolation, we have
\begin{align}
    y(s,t)&=2x(s,t)S(1-S)^2+4x(s,t)S^2\sum_{u<s}x(u,t) \nonumber\\
    & +2x(s,t)\sum_{u\leq M}x(u,t)\sum_{v+w<u+s}x(v,t)x(w,t) \nonumber\\
    & +x(s,t)\sum_{u\leq M}x(u,t) \sum_{v+w=u+s}x(v,t)x(w,t),
    \label{eq:y-x}
\end{align}
for $s=1,...,M$, 
where (by definition) all contributions come from selected edges with an endpoint in a cluster of size $s$. 
The four contributions to the right-hand side of this equation are obtained by considering the cluster sizes of the other endpoint of the selected edge and of the endpoints of the unselected edge, relative to the threshold $M$ and to each other:
(1) the other endpoint of the selected edge is in a cluster of size $>\!M$ and the endpoints of the unselected edge lie in clusters of sizes $\leq\!M$ (first term);
(2) the other endpoint of the selected edge is in a cluster of size $>\!M$ and the unselected edge has one endpoint in a cluster of size $>\!M$ and the other in a cluster of size $<\!s$  (second term);
(3) all endpoints of both edges lie in clusters of size $\leq\!M$, and the sum of the cluster sizes of the unselected edge is strictly smaller than that of the selected edge
(third term);
(4) all endpoints of both edges lie in clusters of size $\leq\!M$ and the sum of the cluster sizes is equal for the selected and unselected edges (fourth term). 
The coupled ordinary differential equations resulting from substituting Eq.~\eqref{eq:y-x}
into Eq.~\eqref{rate_equation}
can be solved numerically to obtain the evolution of $S$ defined in Eq.~\eqref{cluster_size}.

\begin{figure}
    \centering
    \includegraphics[width=1.0\linewidth]{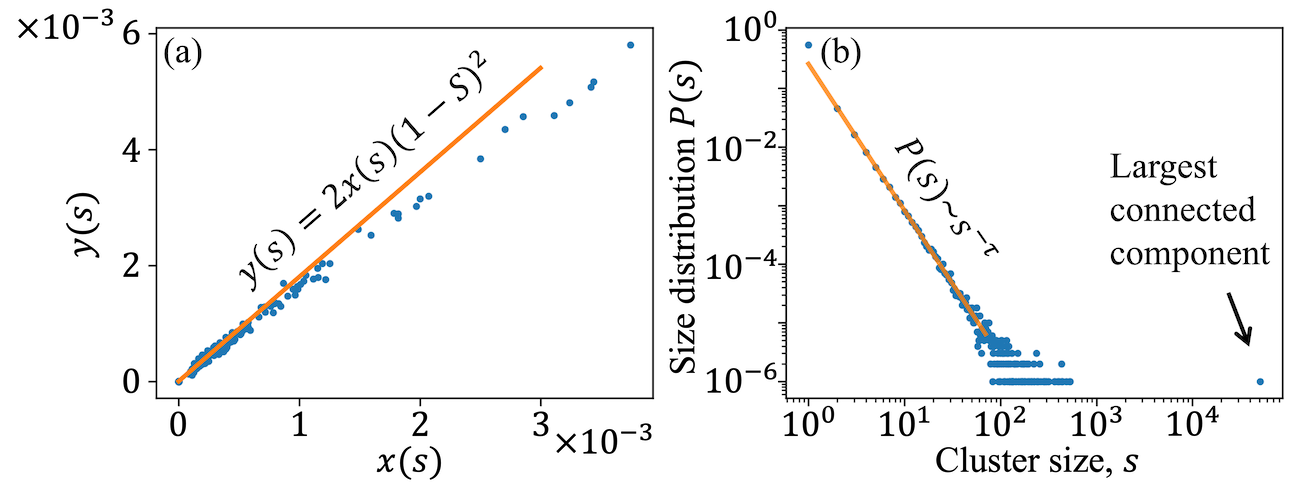}
    \caption{Statistical properties for the PoC protocol at the percolation transition $t_\text{c}$.  
   (a)~Function $y(s)$ vs $x(s)$ for $s>15$ from simulations (blue dots), along with the theoretical prediction in Eq.~\eqref{eq:y-x_sim} (orange line). (b)~Unnormalized cluster-size distribution from simulations (blue dots) and predicted power law with exponent $\tau=2.5$ (orange line). 
    The simulations are for an ER network with $p=0.75$, $\bar{k}=8$, and $N=10^6$. 
    }
    \label{fig:ys_xs}
\end{figure}

Near percolation, the sum $\sum_{u<s}x(u,t)$ in the second term of Eq.~\eqref{eq:y-x} is dominated by small $u$ and can be approximated by $1 - S$ for sufficiently large $s$, even when $s \ll M$. 
The same approximation 
simplifies the third term as a product of three 
independent summations: 
$ 2x(s,t)\sum_{u\leq M}x(u,t) \sum_{v\leq M}x(v,t) \sum_{w\leq M}x(w,t)$.
Finally, the fourth term
is negligible near percolation. With these approximations,  $y(s)$ can be simplified as
\begin{align}
    y(s)&\approx2x(s)S(1\!-\!S)^2+4x(s)S^2(1\!-\!S)+2x(s)(1\!-\!S)^3 \nonumber\\
    &=2x(s)(1-S)(1-S+2S^2) \nonumber\\
    &\approx 2x(s)(1-S)^2, 
    \label{eq:y-x_sim}
\end{align}
where $2S^2$ is neglected since it is expected to be small relative to $S$. This result should be contrasted with the random protocol, where $y(s)=x(s)$. 
Simulation results at the percolation transition with $N=10^6$ show good agreement with Eq.~\eqref{eq:y-x_sim}  [Fig.~\ref{fig:ys_xs}(a)]. 
The transition point $t_\text{c}$ in the simulations is identified as the time at which the derivative $dS(t)/dt$ reaches its maximum, serving as an approximation of the critical point in the limit $N\rightarrow \infty$.

We can now examine the nature of the PoC percolation transition.
From Eqs.~\eqref{rate_equation} and \eqref{eq:y-x_sim},
we obtain $S=\max\{\tilde{S},0\}$, where
 \begin{equation}
     \tilde{S}\approx p\big[-a_0 \Gamma(2-\tau)\big]^{1/(3-\tau)}\big[8(t-t_\text{c})\big]^{(\tau-2)/(3-\tau)}, 
     \label{critical_singularity}
 \end{equation}
$\Gamma$ is the Gamma function, $a_0$ is a normalization factor, and $\tau$ defines the critical distribution $x(s)= a_0s^{1-\tau}$ at $t=t_\text{c}$ (for derivation details, see Appendix~\ref{sec:poc}).
It follows that the (unnormalized) cluster-size distribution $P(s)=s^{-1}x(s)$ scales as $P(s)\sim s^{-\tau}$ at $t=t_\text{c}$.
Under the condition that the scaling function $f(s,t)=x(s,t)/s^{1-\tau}$ is always positive and decays exponentially for $t>t_\text{c}$ to ensure normalizability, the exponent $\tau$ is theoretically constrained to be $5/2$ \cite{dacosta2014solution}. 
Our numerical results for $P(s)$ agree with this prediction [Fig.~\ref{fig:ys_xs}(b)]. Inserting $\tau=5/2$ into Eq.~\eqref{critical_singularity}, we obtain $\tilde S=32\pi p a_0^2 (t-t_\text{c})$ near $t_\text{c}$. 
The percolation transition is of second order since $S$ is continuous and its derivative is zero for $t= t_\text{c}^-$ and strictly positive for $t=t_\text{c}^+$ in the large $N$ limit.
Thus, the PoC protocol induces percolation earlier without altering the order of the transition. A finite-size scaling analysis (Appendix~\ref{sec:finite-analysis}) 
indicates that the rate-equation results capture the qualitative behavior of finite-size networks.

\begin{figure}[b]
    \centering
    \includegraphics[width=0.99\columnwidth] {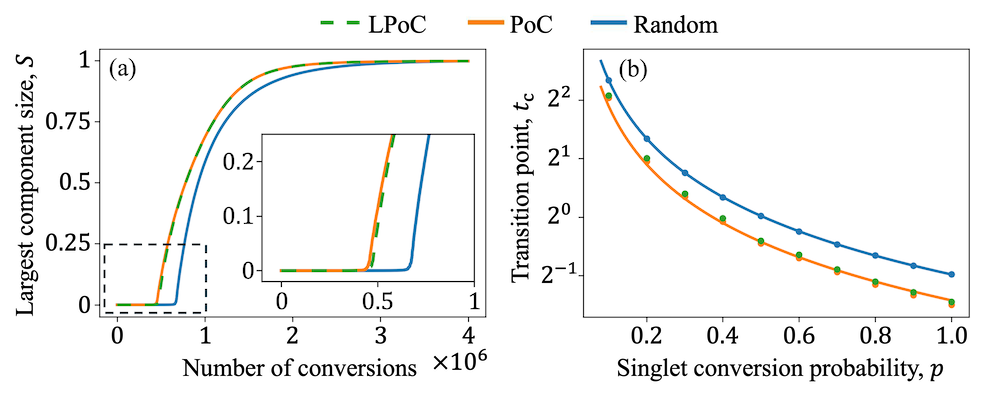}
    \caption{Entanglement percolation for the LPoC protocol. (a)~Evolution of the largest connected cluster size for LPoC, compared with the baseline PoC and random protocols.
    Inset: Zoom-in around the transition.  
    (b)~Entanglement percolation transition point $t_\text{c}$ for LPoC, compared with the PoC and random protocols, as a function of the singlet-conversion probability.
    The curves are the predicted $p^{-1}$ scalings, and the points represent numerical simulations for all three protocols [color-coded as in (a)], with $\bar{k}$ chosen to be larger than $1/p$ to ensure percolation in the ER networks considered.
    In both panels, the unspecified parameters and the number of realizations are the same as in Fig.~\ref{fig:PoC} for $\ell=2$. 
    }
    \label{fig:variance}
\end{figure}

\section{Local implementation}
We now consider an implementation of the PoC protocol based on local information, 
which circumvents its reliance on cluster-size comparisons.  We refer to this formulation as the {\it local power of choice (LPoC) protocol}. Specifically, 
in step (2) of the original protocol, instead of comparing the sum of cluster sizes, we now compare the sum of the  degrees of nodes $a$ and $b$ in the network of converted Bell states with that of nodes $c$ and $d$, and select for singlet conversion the edge with the larger degree sum;
step (1) remains unchanged in the supercritical regime and,
as in the PoC protocol, cluster identity can be neglected in the subcritical regime. 
Protocol implementation based on nearest-neighbor communication, along with the associated communication and memory costs, is further discussed in Appendix~\ref{sec:cost}.

As shown in Fig.~\ref{fig:variance}(a), the LPoC protocol is in excellent agreement with the original PoC protocol
based on
cluster-size
information
(further comparison is presented in SM \cite{SM},  Sec.~S2).
This agreement is significant because, in explosive percolation, the degree-based edge selection yields markedly different results from cluster-based edge selection;
the latter was shown in Ref.~\cite{trevelyan2018degree} for the product rule and is also observed for the summation rule. Moreover, for the PoC protocol, the results remain very similar if the summation rule is replaced by the product rule. However, for the LPoC protocol, the summation rule yields a noticeably earlier percolation transition (SM \cite{SM},  Sec.~S1).

The surprisingly good performance of the LPoC protocol is not limited to the particular singlet-conversion rate illustrated in Fig.~\ref{fig:variance}(a).
To appreciate this, we first note that Eq.~\eqref{rate_equation} is rather general and applies equally well to the random, PoC, and LPoC protocols.
From the linear dependence of $dx(s,t)/dt$ on $p$, we expect the solutions to depend on time through $pt$ for large $s$ and the percolation transition point $t_\text{c}$ to scale as $p^{-1}$ in all three cases. (In the random case, this result is exact and given by $t_\text{c}=1/2p$ \cite{bennaim2005kinetic} per our rate-equation analysis; in the other cases, the prefactor is treated as a fitting parameter.)
Thus, we expect the agreement between LPoC and PoC as well as their advantage over the random protocol to persist for all $p$. 
This is confirmed in our numerical simulations, as shown in Fig.~\ref{fig:variance}(b).

\section{Robustness across networks} 
We now explicitly consider several additional network models with different connectivity patterns. This
includes networks with uniform node degree, scale-free (SF) networks characterized by a power-law degree distribution, and core-periphery networks with a bimodal degree distribution, the latter motivated in part by the relevance of such structures to communication networks. Across all cases, the PoC protocols consistently anticipate the largest component size growth in the supercritical regime, even when the percolation threshold approaches zero.

\begin{figure*}[t]
\centering
    \includegraphics[width=0.58\linewidth]{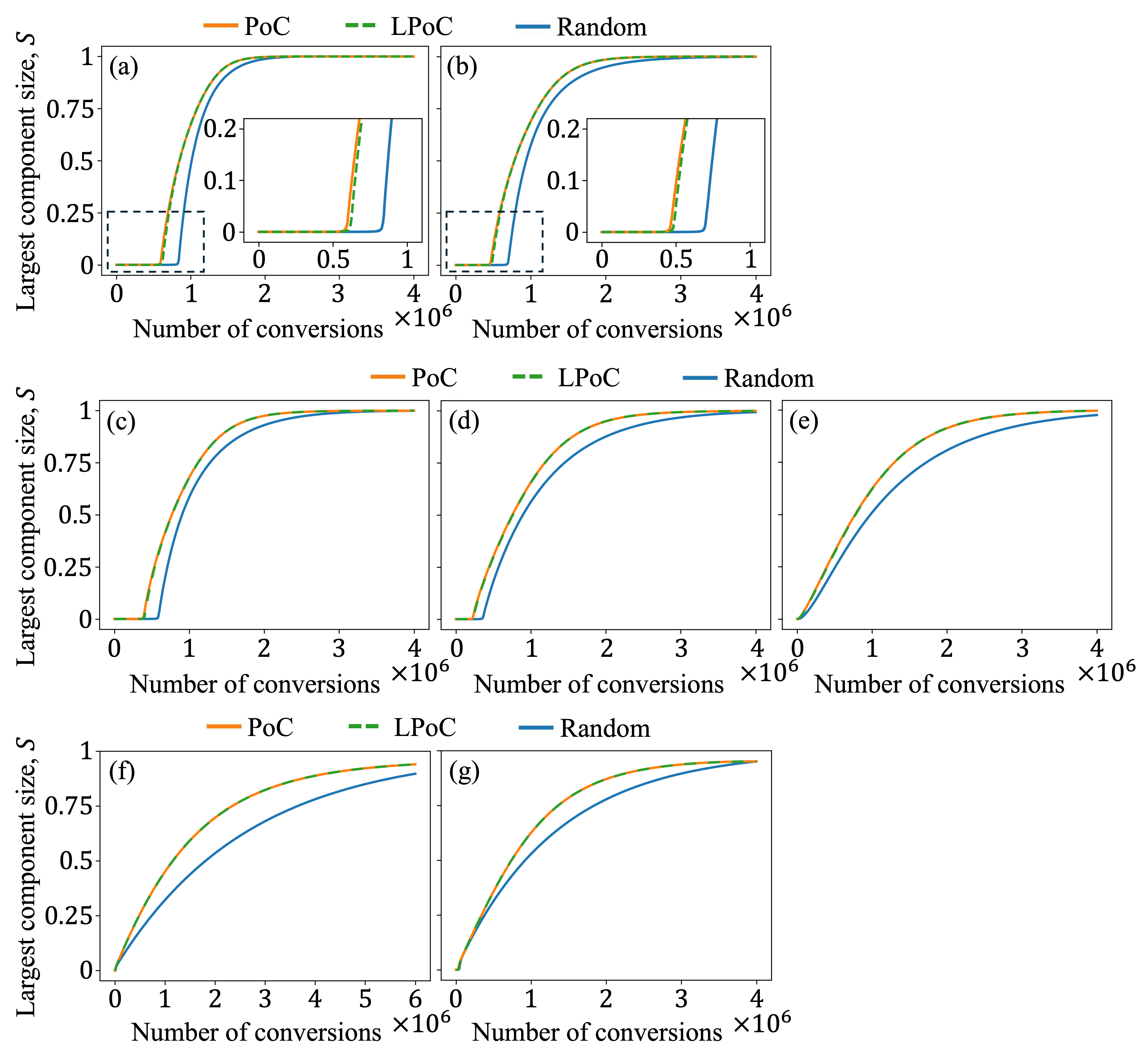}
    \vspace{-0.3cm} \caption{Entanglement percolation for the PoC and LPoC protocols on increasingly heterogeneous networks, with the random protocol shown as a reference. (a) and (b) Random regular networks with degrees $k=5$ (a) and $k=25$ (b), where the curves in (a) are extended beyond the maximum number of possible singlet conversions to facilitate comparison with the other panels. (c)--(e) SF networks with upper degree cutoffs of 25 (c) and 100 (d), and no imposed upper cutoff (e). The scaling exponent for the degree distribution in the SF networks is $\gamma=2.5$,  corresponding to a typical maximum degree $k_\text{max}\approx 5 \times 10^4$ in (e). (f) Core-periphery networks, where the core consists of $10^4$ nodes with an intracore edge probability of $0.08$, and edges between core and peripheral nodes occur with probability $4\times 10^{-4}$. (g) Bipartite limit of a core-periphery network in which the core, of size $10^3$, contains no internal edges, and core-periphery edges are present with probability $4\times 10^{-3}$.
    In all cases, $p=0.75$ and $N=10^6$, and each curve represents an average over ten independent  realizations.}
    \label{fig:scale-free}
\end{figure*}

Specifically, we examine the impact of node degree by first studying 
random regular networks, where all nodes have the same degree $k$. As shown in Figs.~\ref{fig:scale-free}(a) and \ref{fig:scale-free}(b) for $k=5$ and $25$, the advantage of PoC and LPoC over random percolation persists, with the percolation transition threshold decreasing comparably for both protocols as $k$ increases. 
We next turn to SF networks with a degree distribution $P_\text{k}\sim k^{-2.5}$
using different upper bounds on the maximum degree.
As shown in Figs.~\ref{fig:scale-free}(c) and \ref{fig:scale-free}(d), the performance of the proposed protocols remains comparable to that observed for ER networks, even in the presence of substantial degree heterogeneity. In the limiting case of no upper bound on the maximum degree, the percolation threshold approaches zero for all protocols in Fig.~\ref{fig:scale-free}(e), as expected since the second moment of the degree distribution diverges \cite{callaway2000network}.

We further consider a core-periphery model in which a small core, comprising $1\%$ of the nodes, forms an ER network with average degree $k_{\text{core}}\approx 800$. The peripheral nodes, consisting of the remaining $99\%$, are sparsely connected to the core, with each such edge present independently with probability $4\times 10^{-4}$. 
Due to the high connectivity within the core and the other nodes being connected to the core, the percolation transition occurs early even for the random protocol.
However, the PoC and LPoC protocols continue to exhibit a clear advantage over the random protocol in the supercritical regime, anticipating the growth of the largest component size [Fig.~\ref{fig:scale-free}(f)]. We also examine a bipartite limit of a core-periphery model in which the core is small, contains no internal edges, and is strongly connected to the peripheral nodes. Because each core node is connected to a large number of other nodes, the percolation transition occurs early in this case as well. As shown in Fig.~\ref{fig:scale-free}(g), once again the PoC and LPoC protocols accelerate the growth of the largest connected component in the supercritical regime.

In all cases, the minimal average degree for percolation---which is $\bar{k} = 1/p$ in ER networks---is the same for both the random and (L)PoC protocols.
In ER networks, the transition point $t_\text{c}(p)$ does not depend on the average degree, provided that $\bar{k} \ge 1/p$, although the overall percolation curve does, whereas in other networks $t_\text{c}$ generally depends on $\bar{k}$.
Yet, PoC and LPoC reach the transition earlier because their edge-selection strategy preferentially
connects larger clusters together, and they reach the same maximum component size earlier by also preferentially delaying the inclusion of redundant edges.

On the other hand, if  it is possible to make effective use of global information, the protocol can be modified to remove the redundant edges from consideration {\it before} the edge pairs are sampled---a modification that does not alter the transition point but further anticipates the point where the percolation curve saturates
 (SM \cite{SM},  Sec.~S3).
Finally, while we choose to focus on sequential implementations of the PoC protocol and its variants, we note that its advantage over the random strategy persists if singlet conversions on multiple edges are implemented in parallel (SM \cite{SM}, Sec.~S4).

\section{Conclusions}  
Our implementation of an entanglement percolation protocol based on the power-of-choice strategy
provides insight into the resources required for quantum communication. The PoC protocol yields an earlier percolation transition and accelerates the growth of the giant component relative to standard percolation. The results are robust, persisting across the full range of singlet-conversion probabilities and for diverse network topologies, including ER, scale-free, and core-periphery networks. Through a rate-equation analysis, we predict the scaling behavior at the percolation threshold, including the cluster-size distribution, and confirm the second-order nature of the transition.
We emphasize that the PoC protocol admits an implementation based on local information with no appreciable loss of performance, which is both highly desirable and surprising, since analogous local implementations yield markedly different results from their global counterparts in (classical) explosive percolation~\cite{trevelyan2018degree}.
The proposed approach is amenable to decentralized implementation, making it relevant to ongoing efforts toward large-scale quantum communication.

\let\oldaddcontentsline\addcontentsline
\renewcommand{\addcontentsline}[3]{}

\smallskip

\section*{Acknowledgment}

This work was supported by ARO MURI Grant No.~W911NF-21-1-0325. 

\section*{Data availability}
The data that support the findings of this article were generated by numerical simulations and are available from the authors upon reasonable request. 


{\appendix

\section{Nature of the Power-of-Choice Percolation Transition}
\label{sec:poc}
To examine the nature of the PoC percolation transition, we adapt a method used by da Costa \textit{et al.} \cite{dacosta2015inverting, dacosta2010explosive}. We first define  generating functions for $x(s,t)$ and $y(s,t)$:
\begin{align}
    \rho(z,t)=\sum_{s}z^s x(s,t), \\
    \sigma(z,t) = \sum_{s}z^s y(s,t).
\end{align}
Multiplying both sides of Eq.~\eqref{rate_equation} (main text) by $z^s$ and summing over $s$,
we obtain
\begin{equation}
    \frac{\partial \rho(z,t)}{\partial t}=2p\big[\sigma(z,t)-1\big]\frac{\partial \sigma(z,t)}{\partial \ln z}. 
    \label{eq:rho}
\end{equation}
For $z$ close to $1$, the functions $\sigma(z)$ and $\rho(z)$ can be related as
\begin{align}
    \sigma(z)
    &=\sum_s y(s) + \sum_s y(s)(z^s-1) \nonumber\\
    &\approx (1-S)^2 + 2(1-S)^2\sum_s x(s)(z^s-1) \nonumber\\
    &=2(1-S)^2 \rho(z)+(1-S)^2(2S-1),
    \label{eq:sigma}
\end{align}
where we used Eq.~\eqref{eq:y-x_sim} (main text) and the normalization conditions $\sum_s x(s)=1-S$ and $\sum_s y(s)=(1-S)^2$. Substituting Eq.~\eqref{eq:sigma} into Eq.~\eqref{eq:rho} yields
\begin{equation}
    \frac{\partial \rho}{\partial t} \approx \; 4p(1-S)^2\big[(1-S)^2(2\rho+1)-2(1-S)^3-1\big]\frac{\partial \rho}{\partial \ln z}. 
\end{equation}
Applying a hodograph transformation and setting $z=1$,
we obtain $S\approx \max\{\tilde{S},0\}$, where
\begin{equation}
    \tilde{S}\approx p\big[-a_0 \Gamma(2-\tau)\big]^{1/(3-\tau)}\big[8(t-t_\text{c})\big]^{(\tau-2)/(3-\tau)}. 
    \label{critical_singularity}
\end{equation}
Here,
$\Gamma$ is the Gamma function, $a_0$ is a normalization factor, and $\tau$ defines the critical distribution $x(s)= a_0s^{1-\tau}$ at $t=t_\text{c}$. As explained in the main text, it follows from Eq.~\eqref{critical_singularity} that the PoC protocol gives rise to a second-order percolation transition.

\setcounter{figure}{0}
\renewcommand{\thefigure}{B\arabic{figure}}

\section{Finite-Size Scaling Analysis}
\label{sec:finite-analysis}

In finite networks, the percolation transition is rounded, and the critical point is not sharply defined due to finite-size fluctuations. Finite-size scaling analysis allows one to extract the thermodynamic-limit behavior from finite systems
\cite{newman1999monte}. Thus,
to characterize the transition more thoroughly, we take the order parameter to be $S(t, N)$, the normalized size of the largest connected component, where we now explicitly indicate its dependence on the normalized time step $t$ and network size $N$.
Near the critical point $t_c$, the order parameter is expected to obey the finite-size scaling ansatz 
\begin{equation}
S(t,N)=N^{-\beta/\nu} \mathcal{F}\big((t-t_c)N^{1/\nu}\big),
\label{eq:fss}
\end{equation}
where $\beta$ and $\nu$ are critical exponents and $\mathcal{F}$ is a universal scaling function independent of $N$. This ansatz captures how the singular behavior at criticality is rounded in finite systems for both the random and PoC protocols. 

Numerically, the transition point $t_c(N)$ is identified as the time of maximal variation of $S(t, N)$. At the critical point $t_c(N)$, the order parameter scales as $S(t_c,N)\sim N^{-\beta/\nu}$; equivalently,
\begin{equation}
    S(t_c,N)N\sim N^{1-\beta/\nu}.
\end{equation} 
Further, the width of the transition scales as $\Delta t\sim N^{-1/\nu}$, which implies
\begin{equation}
    \Delta tN\sim N^{1-1/\nu}.
    \label{eq:width}
\end{equation}
To extract the critical exponents and validate the scaling function, we simulate networks of different sizes, ranging from $N=3.125\times 10^4$ to $N=10^6$. As shown in Fig.~\ref{fig:finite}(a), both the random and PoC protocols exhibit a linear relationship between the logarithm of $SN$ and the logarithm of  $N$ at their respective critical points, where the slope of the fitted lines approximates $1-\beta/\nu$. Furthermore, Figs.~\ref{fig:finite}(b) and \ref{fig:finite}(c) show the collapse of $S(t,N)N^{\beta/\nu}$ as a function of $(t-t_c)N^{1/\nu}$ onto a single universal function $\mathcal{F}$ for different $N$ [using $\beta/\nu$ obtained from Fig.~\ref{fig:finite}(a) and a suitable value of $\nu$ for each protocol]. This collapse provides evidence that the percolation transition in our system is governed by well-defined critical exponents and supports the validity of the finite-size scaling framework. 
As shown in Fig.~\ref{fig:finite}(d), the scaling in Eq.~(\ref{eq:width}) for the same $\nu$ agrees well with the numerically estimated percolation transition width, given by the difference between the values of $t$ at which $S(t,N)$ reaches $S(t_c,N)/2$ and $2S(t_c,N)$.

\begin{figure}[h]
    \centering
    \includegraphics[width=0.99\linewidth]{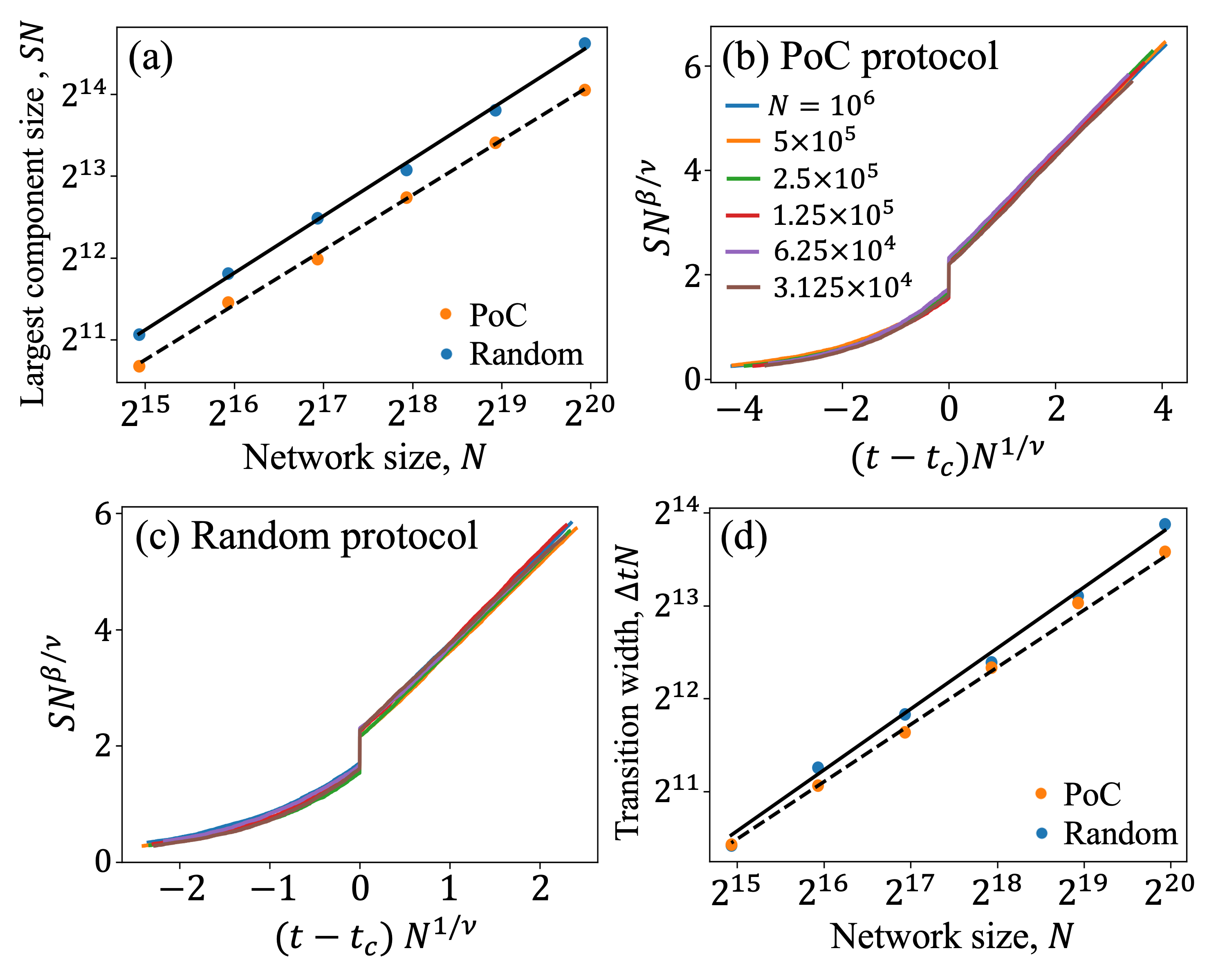}
    \caption{Finite-size scaling analysis for the random and PoC protocols. (a) Linear fits of $\log_2 SN$ vs $\log_2 N$, yielding slopes of        $0.67$ and $0.70$ for the  PoC and random protocols, respectively. (b) and (c) Data collapse of the scaling function $\mathcal{F}$ defined in Eq.~\eqref{eq:fss}, with the curves in both panels following the same color code. For the PoC protocol (b), $\beta=0.86$ and $\nu=2.6$. For the random protocol (c), $\beta=0.88$ and $\nu=2.9$. 
    (d) Linear dependence of $\log_2 \Delta tN$ on $\log_2 N$, with the lines indicating the predictions for $\nu$ from (b) and (c). The simulations are based on ER networks with $p=0.75$ and $\bar{k}=8$, and each data point represents an average over $200$ independent realizations.
    }
    \label{fig:finite}
\end{figure}

\section{Implementation Details and Cost of the Different Protocols}
\label{sec:cost}
In all the protocols discussed in this work, including the random baseline, each node is required to maintain a record of its local entanglement resources, including the set of entangled states available to it for future singlet conversion and those already successfully converted. 
This accounting is inherently local and is updated whenever an entangled state is selected for singlet conversion.

For the PoC protocol, nodes need to additionally maintain a record of the connected component to which they belong. Specifically, each node stores its cluster index and the associated cluster size. At initialization, every node is assigned a unique index, and all clusters have size one. During the percolation process, when $\ell$ edges are randomly selected, the pair of end nodes of each edge first compare their cluster indices to determine whether they already belong to the same connected component. If singlet conversion of an edge connecting two different clusters is successful, the clusters are merged. The merge requires the end nodes to select one of the existing indices as the new cluster index, compute the updated cluster size as the sum of the original cluster sizes, and propagate this information across the new cluster via local node-to-node communication through Bell-connected nodes. 
Note that the fraction of redundant edges is negligible before the percolation transition; hence, tracking cluster sizes alone is sufficient, while propagation of cluster indices becomes necessary only after the transition.

For LPoC, the process is simpler in two ways. First, each node keeps a local record of its entanglement resources, and thus of its degree in the Bell-state network. Second, no information propagation is required in the subcritical regime due to the negligible fraction of redundant edges. Consequently, this protocol does not require information propagation across the new cluster prior to the transition point, and only requires propagation of the cluster index after the transition.

The enhanced power-of-choice protocol discussed in SM \cite{SM} Sec.~S3
can be implemented via nearest-neighbor propagation of cluster size (and cluster index after the transition) following each cluster merge, as in the PoC case.
Additional information is required to exclude redundant partially entangled states before sampling random edge pairs at each time step. These states can be identified and flagged during this propagation. Even though this introduces additional communication and memory costs, the protocol still relies only on node-to-node communication.

Therefore, compared with the random protocol, the protocols in this study involve additional memory and communication costs, which are nevertheless scalable. For a node $i$, we denote the degree in the original network by $k_i$, the degree in the Bell-state network by $k_i^{(\text{B})}$, and the degree in the network of remaining unconverted states by $k_i'$, where $k_i' \le k_i - k_i^{(\text{B})}$ at all times. We measure memory and communication costs in terms of the number of stored and transmitted integers, respectively, each bounded by $N$.
For each node, the memory cost of storing the cluster index and cluster size is $\mathcal{O}(1)$; the cost of storing redundant edges is at most $\mathcal{O}(k_i')$. Moreover, for each node in a merging cluster, the per-merge communication cost of propagating the cluster index and size is $\mathcal{O}(k_i^{(\text{B})})$, while the cost of identifying redundant edges is $\mathcal{O}(k_i')$. The random protocol itself requires up to $\mathcal{O}(k_i)$ memory per node to store local entanglement resources.

}

\let\addcontentsline\oldaddcontentsline

\onecolumngrid

\newpage

\startcontents[sm]

\renewcommand{\thepage}{\arabic{page}}
\setcounter{page}{1}

\renewcommand{\thefigure}{S\arabic{figure}}
\setcounter{figure}{0}

\setcounter{secnumdepth}{3}
\renewcommand{\thesection}{S\arabic{section}}
\setcounter{section}{0}

\renewcommand{\theequation}{S\arabic{equation}}
\setcounter{equation}{0}

{\centering
{\Large\sc{Supplemental Material}}\\
\vspace{1mm}
for \\
\vspace{1mm}
{\Large Entanglement Percolation Enhanced by the Power of Choice}\\
\vspace{1mm}
Yanxuan Shao and Adilson E. Motter\\
}

\bigskip
\bigskip

\titlecontents{section}
  [1.5em]
  {\vspace{4pt}}
  {\contentsmargin{0em}\thecontentslabel\hspace{1em}}  
  {\hspace{0em}}
  {\normalfont\hfill\thecontentspage}

\printcontents[sm]{}{1}

\bigskip
\bigskip
\bigskip

\section{Protocol Using Product vs.\ Summation Rules} \label{sec:production_and_summation}
\begin{figure}[h]
    \centering
    \includegraphics[width=0.6\linewidth]{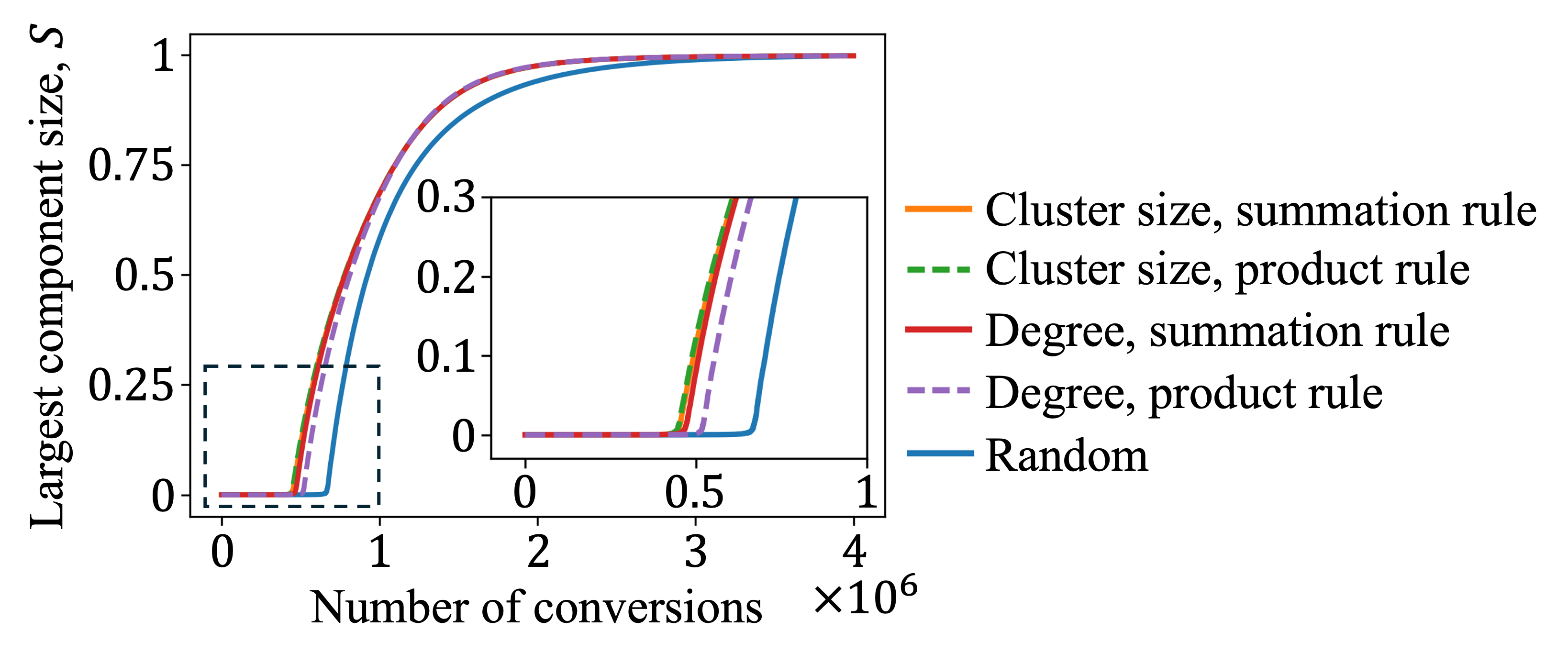}
    \caption{Entanglement percolation for the PoC protocol using the summation rule and the product rule, for selection criteria based on cluster size and node degree. The random protocol is shown as a reference.
    The parameters are $p=0.75$, $\bar{k}=8$, and $N=10^6$, where each percolation curve is an average over $10$ realizations.}
    \label{fig:prod_sum}
\end{figure}

In the main text, we implement the PoC protocol using a {\it summation rule}, where edges are selected for singlet conversion based on the sum of the sizes of the clusters they connect or of the sum of the degrees of the end nodes. In the study of explosive percolation in classical networks, a {\it product rule} has been considered, where the edge selection is based on the product (rather than the sum) of cluster sizes and node degrees, respectively. As shown in Fig.~\ref{fig:prod_sum}, 
for cluster sizes, the results are nearly indistinguishable with either the summation or the product rule; for node degrees, the summation rule leads to slightly earlier percolation than the product rule. Throughout the paper, we chose to focus on the summation rule in order to have a formulation more amenable to mathematical treatment via rate equations.

\newpage
\section{Node Degree vs. Cluster Size Rules}
\label{sec:clusterSize_degree}
In the main text, we show that the percolation process under the LPoC protocol aligns closely with the original PoC protocol. To quantify the relationship between a node's degree and the size of its associated connected component, we first examine the percolation prediction from the generating function framework for configuration model networks. 

Let $p(k)$ denote the degree distribution of the network, and define its generating function as $G_0(x)=\sum_k p(k)x^k$. The distribution of outgoing edges is then generated by $G_1(x)=\sum_k \frac{kp(k)}{\langle k\rangle}x^{k-1}$. Let $H_1(x)$ be the generating function for the size distribution of the cluster reached by following a random edge, which satisfies the self-consistent equation $H_1(x)=xG_1[H_1(x)]$. Assuming a tree-like structure, the generating function for the component size of a node of degree $k$ is $H_k(x)=x[H_1(x)]^k$. The mean component size for a node of degree $k$ is then given by $\langle s\rangle_k=\frac{dH_k(x)}{dx}|_{x=1}=1+kH_1'(1)=1+\frac{k}{1-G'_1(1)}$. This result shows that, below the percolation transition, the expected cluster size is linearly dependent on the node's degree for the networks specified by $p(k)$. 

\begin{figure}[h]
    \centering
    \includegraphics[width=0.65\linewidth]{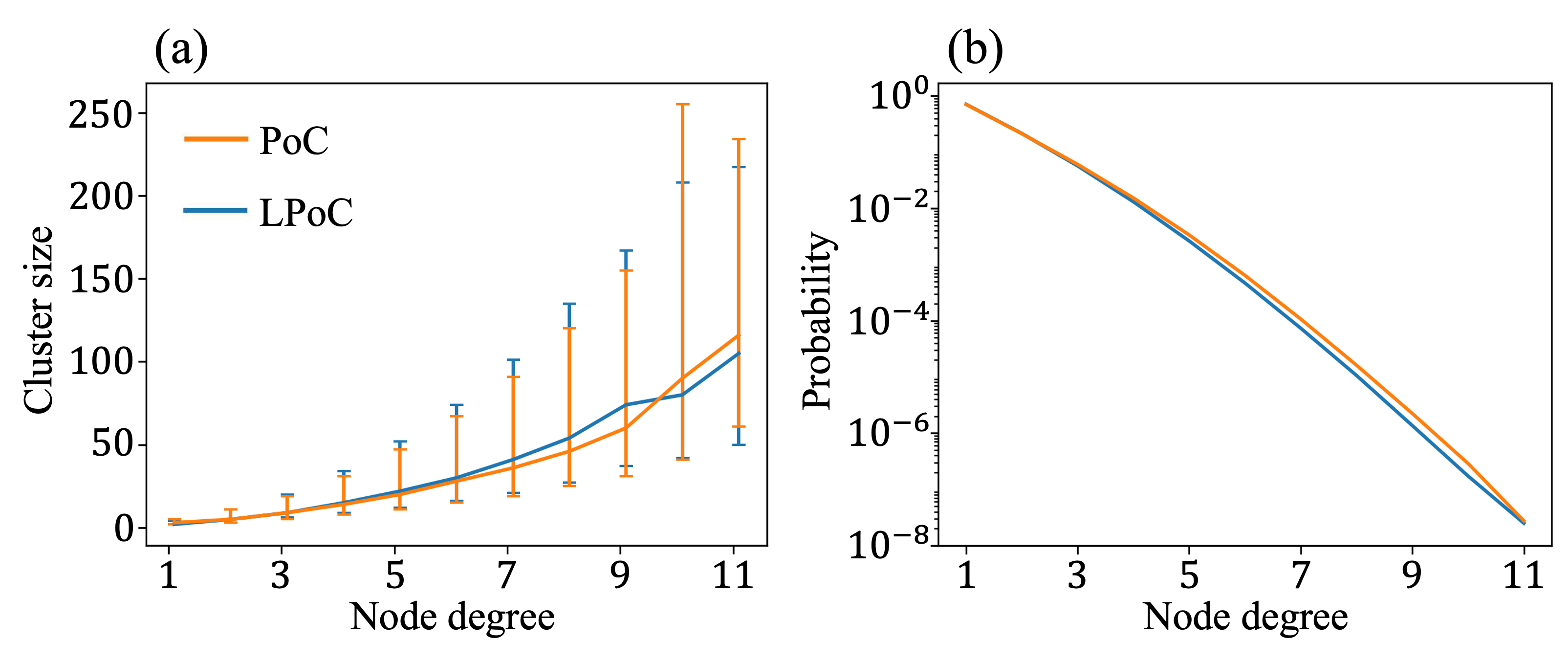}
    \caption{Cluster size versus node degree in the subcritical regime. (a) Median cluster size as a function of node degree for the PoC and LPoC protocols. The error bars indicate the interquartile range. (b) Degree distribution of nodes at the ends of randomly chosen edges, color coded as in (a). In both panels, the cluster size and node degree are defined with respect to the maximally entangled network, consistent with the rules of the PoC and LPoC protocols. The parameters are $p=0.75$, $\bar{k}=8$, and $N=10^6$, where each curve is an average over $1000$ realizations.}
    \label{fig:cluster_degree}
\end{figure}

Our numerical results show an analogous dependence for the PoC and LPoC protocols prior to the percolation transition. Focusing on the cumulative distribution over the entire subcritical phase and using the median to improve statistical stability, we observe a monotonic relationship between cluster size and the degree of the endpoint nodes of randomly chosen edges [Fig.~\ref{fig:cluster_degree}(a)]. This relationship and the associated degree distributions [Fig.~\ref{fig:cluster_degree}(b)], which are comparable across both protocols, suggest that similar percolation dynamics emerge whether the protocol is based on cluster size or node degree.

\newpage
\section{Enhanced Power-of-Choice Protocol} \label{sec:epoc}
\begin{figure}[h]
    \centering
\includegraphics[width=0.4\linewidth]{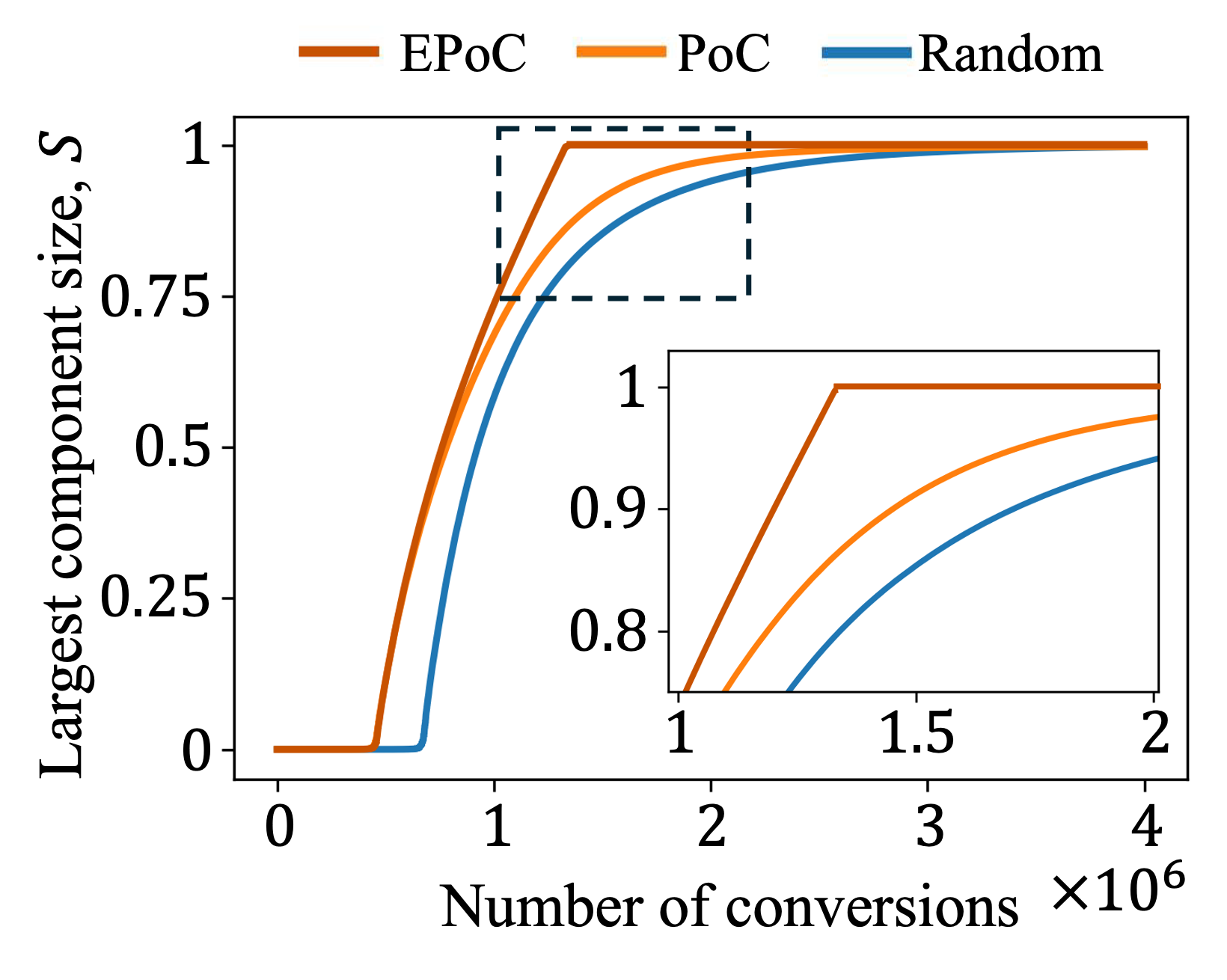}
\caption{Entanglement percolation for the enhanced PoC protocol, with the baseline PoC and random protocols shown as references.
    Inset: Zoom-in as the percolating network approaches full connectivity.
    The parameters are $p=0.75$, $\bar{k}=8$, and $N=10^6$, where each percolation curve is an average over $10$ realizations.     
}
    \label{fig:epoc}
\end{figure}

We have also considered the variant of the PoC protocol, which we refer to as the {\it enhanced power-of-choice (EPoC) protocol}, in which all partially entangled edges connecting nodes from the same clusters are treated as redundant and removed from consideration before edge pairs are randomly sampled. As shown in Fig.~\ref{fig:epoc},
the EPoC protocol exhibits the same percolation point as the PoC protocol, since the fraction of redundant edges is negligible before the transition. 
Above percolation, however, it accelerates the growth of $S$ toward saturation, because each interaction consists only of step (2), where the comparison necessarily involves edges linking different clusters, thereby bypassing the increasing number of redundant edges.

\newpage
\section{Parallel Power-of-Choice Protocol} \label{sec:ppoc}
In the main text, we consider {\it sequential} formulations of the PoC protocol in which at most one singlet conversion is performed at each time step, and the outcome of that step is accounted for in all subsequent ones. Here, we examine a variant of the protocol in which all nodes may perform singlet conversion on one of their edges in {\it parallel}.
This formulation starts with the same $N$-node network of partial entanglements, but now at each time step, each node selects one of its remaining partially entangled edges to perform singlet conversion. 
(The nodes are assumed to communicate classically to coordinate, ensuring that only one of them performs the local operations associated with the singlet conversion.)
Specifically, each node with two or more unconverted edges selects one out of two randomly selected edges that, if successfully converted, would maximize the resulting cluster size; nodes with only one unconverted edge always implement singlet conversion on that edge.
We refer to this variant of the PoC protocol as the {\it parallel power-of-choice (PPoC) protocol}.

\begin{figure}[h]
    \centering
    \includegraphics[width=0.4\linewidth]{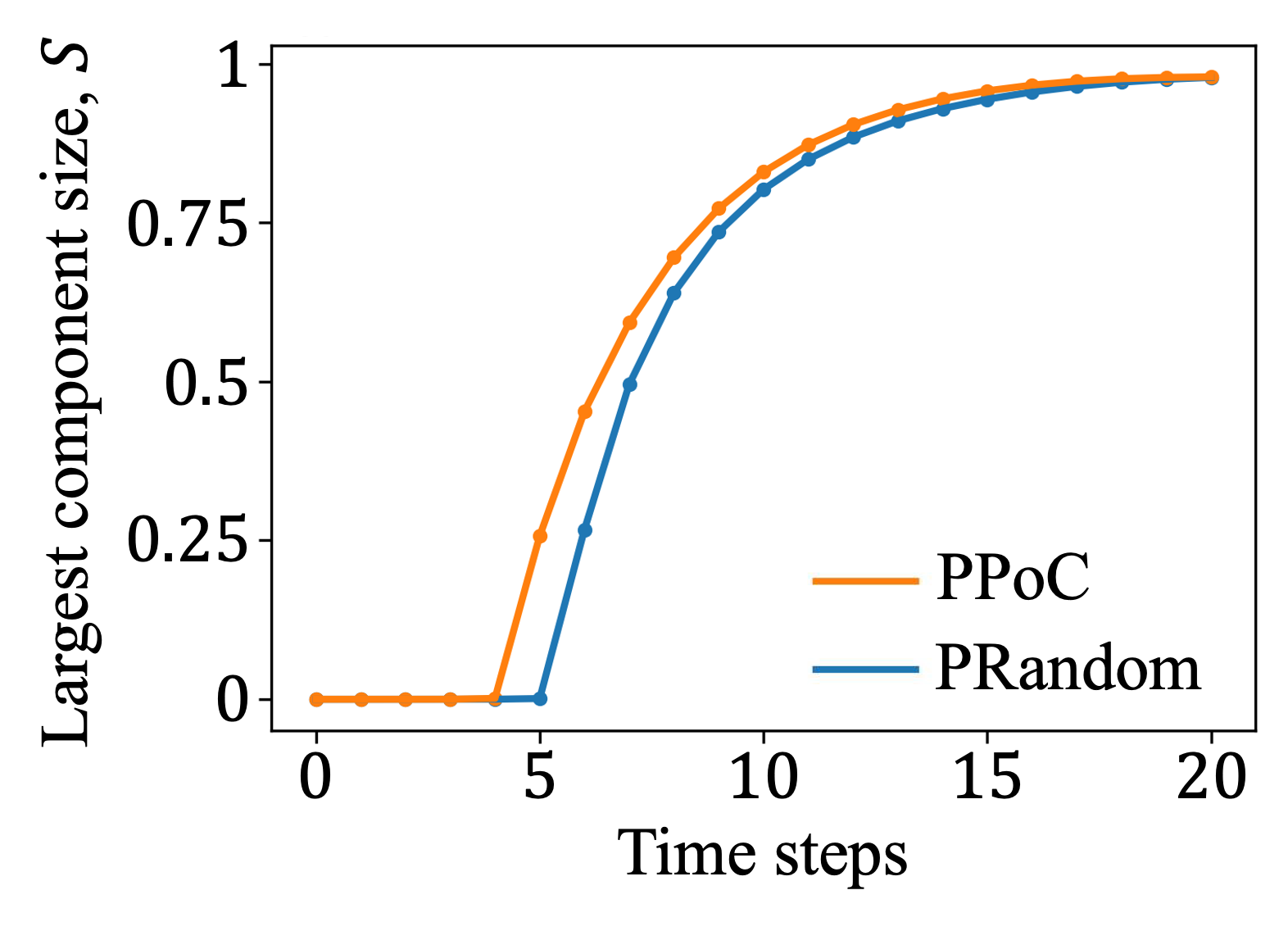}
    \caption{Entanglement percolation for the parallel PoC protocol, shown together with the parallel implementation of the random protocol for reference. The parameters are $p=0.1$, $\bar{k}=40$, and $N=10^6$, where each percolation curve represents an average over $10$ realizations.}
    \label{fig:ppoc}
\end{figure}

As shown in Fig.~\ref{fig:ppoc},
the PPoC protocol reaches the percolation transition within just a few time steps. This is the case because the process now involves up to $N$ singlet conversions per time step.
The total number of operations, however, is generally not smaller than in the baseline PoC protocol (and can in fact be larger) if all parameters are the same. We note, however, that we intentionally chose a smaller $p$ and larger $\bar{k}$ in Fig.~\ref{fig:ppoc} than in the main text in order to have enough time steps to more clearly contrast it with the random protocol. To
ensure a fair comparison, we consider a parallel implementation of the random protocol (PRandom) where, at each time step, an edge of each node is selected for singlet conversion.

\end{document}